\PassOptionsToPackage{pdfpagelabels=false}{hyperref}
\RequirePackage{fix-cm}
\documentclass[fleqn,usenatbib]{mnras}

\usepackage[T1]{fontenc}
\usepackage{ae,aecompl}

\usepackage{graphicx}	
\usepackage{amsmath}	
\usepackage{amssymb}	
\usepackage{epstopdf} 
\usepackage[para]{threeparttable} 
\usepackage{subfig}
\usepackage{pstricks}
\usepackage{float}
\usepackage{latexsym}
\usepackage{hyperref}

\title[Activity of (2060) Chiron possibly caused by impacts?]{Activity of (2060) Chiron possibly caused by impacts?}

\author[S. Cikota et al.]
{Stefan Cikota$^{1,2}$\thanks{E-mail: stefan.cikota@fer.hr},
Estela Fern\'{a}ndez-Valenzuela$^{3}$,
Jose Luis Ortiz$^{3}$,
Nicol\'{a}s Morales$^{3}$,
\newauthor Ren\'{e} Duffard$^{3}$,
Jesus Aceituno$^{4}$,
Aleksandar Cikota$^{5}$,
Pablo Santos-Sanz$^{3}$
\\
$^{1}$University of Zagreb, Faculty of Electrical Engineering and Computing, Department of Applied Physics, Unska 3, 10000 Zagreb, Croatia \\
$^{2}$Ru{\dj}er Bo\v{s}kovi\'{c} Institute, Bijeni\v{c}ka cesta 54, 10000 Zagreb, Croatia \\
$^{3}$Instituto de Astrof\'{i}sica de Andaluc\'{i}a - CSIC, Apt 3004, 18008 Granada, Spain \\
$^{4}$Centro Astron\'omico Hispano Alem\'an de Calar Alto (CSIC-MPG), C/ Jes\'us Durb\'an Rem\'on 2-2, E-4004 Almer\'ia, Spain. \\
$^{5}$European Southern Observatory, Karl-Schwarzschild-Str. 2, 85748 Garching b. M\"{u}nchen, Germany  
}

\date{Accepted XXX. Received YYY; in original form ZZZ}

\pubyear{2017}

\begin{document}
\label{firstpage}
\pagerange{\pageref{firstpage}--\pageref{lastpage}}
\maketitle

\begin{abstract}
The centaur 95P/(2060) Chiron is showing comet-like activity since its discovery, but the mass-loss mechanisms triggering its activity remained unexplained.
Although the collision rates in the centaur region are expected to be very low, and impacts are thought not to be responsible for the mass-loss, since the recent indications that Chiron might possess a ring similar to Chariklo's, and assuming that there is debris orbiting around, the impact triggered mass-loss mechanism should not be excluded as a possible cause of its activity. 
From time series observations collected on Calar Alto Observatory in Spain between 2014 and 2016, we found that the photometric scatter in Chiron's data is larger than a control star's scatter, indicating a possible microactivity, possibly caused by debris falling back to Chiron's surface and lifting small clouds of material. 
We also present rotational light curves, and measurements of Chiron's absolute magnitudes, that are consistent with the models supporting the presumption that Chiron possesses rings.
By co-adding the images acquired in 2015, we have detected a $ \sim $5 arcsec long tail, showing a surface brightness of 25.3 mag(V)/arcsec$^{2}$.
\end{abstract}



\section{Introduction}

Chiron (1977 UB), officially designated as both, minor planet (2060) Chiron and comet 96P/Chiron, was discovered on 1 November 1977 by astronomer Charles T. Kowal. Chiron is determined to be a taxonomic class C object \citep{Hartmannetal1990}, with a diameter estimated to 215.6 $ \pm $ 9.9 km \citep{Fornasieretal2013}. It was the first of a new class of objects discovered in our Solar System, the so-called ``centaurs''.

Centaurs are objects which orbit between Jupiter and Neptune. They have probably been scattered from the Kuiper Belt in recent solar system history \citep[e.g.][]{SaridPrialnik2009}. \cite{OikawaEverhart1979} have found that Chiron's orbit is not stable, suggesting short dynamical lifetimes.
Numerical simulation of orbits like that of Chiron suggest that Chiron may have been a short-period comet in the past, and will probably become one in the future \citep{HahnBailey1990}.

Shortly after its discovery a cometary nature was suspected \citep{Kowaletal1979}. Those speculations were confirmed in 1988 by VRIJHK photometric measurements \citep{Tholenetal1988} and in 1989 by CCD images showing a comet-like coma \citep{MeechBelton1990}. Impulsive brightenings on a timescale of hours have been detected by \cite{LuuJewitt1990}, but also on time scales of a few to several years \citep{Busetal1991}.

Photographic photometry of (2060) Chiron was carried out using archival plates taken between 1969 and 1977, showing extensive intervals of brightening when Chiron was near aphelion \cite{Busetal2001}. Those anomalous brightness variations were also observed between 1980-1983 \citep{Hartmannetal1990}, and near the end of the 1980s when Chiron approached its perihelion \citep{Busetal2001}.
Even when it was near its historical brightness minima, in post-perihelion observations obtained between 1996 and 1998, a coma was clearly detected \citep{SilvaCellone2001}.

The presence of dust in Chiron's inner coma, including what were supposed to be comet-like dust jets from the nucleus \citep{Elliotetal1995, Busetal1996}, and symmetric jet-like features \citep{Ruprechtetal2015}, were detected by occultation events in 1993, 1994 and 2011 respectively.

The mass-loss mechanisms on active asteroids remain often unexplored. Sublimation of water ice, and thermal processes are often suspected to be possible reasons for ejecting material from the asteroids' surfaces \citep{Jewitt2012}. 
Spectroscopic indications for presence of water ice on Chiron has already been detected in the near-infrared spectra \citep{Luuetal2000, RomonMartinetal2003}. But by reviewing the history of Chiron's activity, it is easily noticeable that its activity is not related to the heliocentric distance, indicating that the mass loss is probably not caused by ice-sublimating mechanisms.

Impacts are a further possible trigger for mass-loss. Collisional processes occur frequently between the small bodies in the main belt. Apart from the collisional origin of asteroid families in the Main Belt \citep[e.g.][]{Tedesco1979, Zappalaetal2002}, the best witnesses of asteroid collisions are outbursts of P/2010 A2 (LINEAR) \citep[e.g.][]{MorenoA2, Snodgrass2010} and (596) Scheila \citep[e.g.][]{Moreno596, Jewitt2011}, both in 2010, which have proven that impacts between asteroids can create large debris clouds of dust- and gravel-sized particles.
Unlike the collision rates in the main belt, the collision rates in the Edgeworth-Kuiper Belt and centaur region are expected to be much lower. \cite{DurdaStern2000} estimated cratering collisions of 1 km radius comets onto Chiron-size targets to occur every $ \sim $ 60 Gyr. Consequently the mass-loss mechanism is thought not to be caused by impacts.

After the recently discovered rings of the centaur (10199) Chariklo \citep{BragaRibasetal2014}, the secondary dimmings in occultation light curves of Chiron formerly interpreted as comet-like dust jets \citep{Ruprechtetal2015}, were interpreted by \cite{Ortizetal2015} to be rings with similar properties to those of Chariklo \citep{Duffardetal2014}.

Assuming that there is debris around Chiron, it is well possible that some of this debris may be continually falling and impacting on its surface. Consequently, although the collision rates in the Centaur region are expected to be very low, a dust release mechanism triggered by impacts should not be excluded as a possible cause for its activity.

In this paper we derived Chiron's light curves, searched for comet-like activity signs and, particularly, aimed to study something which we find an interesting possibility: whether the scatter in Chiron's photometric data is larger than the scatter of a star of identical magnitude as Chiron. Showing that Chiron has continuous micro-brightness changes would indicate an activity possibly caused by debris falling back to the surface and lifting small clouds of material.

\section{Observations and data reduction}

We carried out three observation campaigns from 2014 to 2016, using three different telescopes at Calar Alto Observatory (CAHA) in Almer\'ia, Spain (Table \ref{ObsTab}). 

The first observation campaign was carried out in order to obtain Chiron's rotational light curve. This campaign is composed of three runs. The first run took place on 2014, July 18, 19 using the MOSCA focal reducer, installed at the Richey-Chretien Focus of the 3.5 m telescope. This device has a  field of view (FoV) of 11' x 11' and a pixel scale of 0.33"pixel$^{-1}$ (unbinned). The images were obtained using $R$ band in the Bessell's filters system and in 2 x 2 binning mode. The second and third runs were carried out on 2014 July 23, 24 and October 27, 28 with DLR-MKIII CCD camera of the 1.23 m telescope. The FoV and the image scale of the instrument are 21.5' x 21.5' and 0.314"pixel$^{-1}$ (unbinned), respectively. We took the images in 2 x 2 binning mode and without filter to maximize the signal noise ratio (SNR). A total of 136 images were taken during the whole campaign.

The second observation campaign was executed on 2015, September 11 - 14 in order to obtain Chiron's rotational light curve, determine Chiron's absolute magnitude, and search for comet-like activity signs. We used the Focal Reducer and Faint Object Spectrograph (CAFOS) instrument. This CCD detector, located at the 2.2 m telescope, has a FoV of 16', and an image scale of 0.53"pixel$^{-1}$ (unbinned). The images were taken in the 2 x 2 binning mode, using an $R$ band Bessell's filter. A total of 124 images were analyzed in this campaign. The images taken on September 14 were not used in the rotational light curve because Chiron's flux was particularly contaminated by the flux of other objects.

Finally, the third observation campaign took place on 2016, September 2 in order to obtain a further measurement of Chiron's absolute magnitude. We used the same instrumentation as in the second campaign. Images were taken using $R$ and $V$ bands in the Bessell's filters system and in 2 x 2 binning mode. A total of 6 images of Chiron were taken during this last campaign.

\begin{table*}
\centering
\caption{Journal of Observations for Chiron at the Calar Alto Observatory. Filters are based on the Bessell system. Abbreviations are defined as follows: exposure time ($t_{exp}$), heliocentric distance (R), geocentric distance ($\Delta$) and number of acquired images (N).}
\label{ObsTab}
\begin{tabular}{cccccccc} 
\hline
Date & Telescope & Instrument & Filter         & $t_{exp}$             &  R      & $\Delta$    & N \\
	 &           &            &                & (seconds)             & (AU)    & (AU)        &   \\
\hline
2014 Jul 18 & 3.5 m      & MOSCA      & R          & 300              & 17.878 &  17.238 & 19\\
2014 Jul 19 & 3.5 m      & MOSCA      & R          & 300              & 17.878 &  17.225 &14 \\
2014 Jul 23 & 1.23 m     & DLR-MKIII  & Clear      & 300              & 17.881 &  17.178 & 8\\
2014 Jul 24 & 1.23 m     & DLR-MKIII  & Clear      & 300              & 17.882 &  17.167 & 27\\
2014 Oct 27 & 1.23 m     & DLR-MKIII  & Clear      & 240              & 17.956 &  17.303 & 28\\
2014 Oct 28 & 1.23 m     & DLR-MKIII  & Clear      & 240              & 17.956 &  17.310 & 40\\
2015 Sep 11 & 2.2 m      &    CAFOS   &  R         & 300              & 18.181 &  17.178 & 46 \\
2015 Sep 13 & 2.2 m      &    CAFOS   &  R         & 300              & 18.182 &  17.179 & 70 \\
2015 Sep 14 & 2.2 m      &    CAFOS   &  R         & 300              & 18.184 &  17.182 & 8  \\
2016 Sep 02 & 2.2 m      &    CAFOS   &  R, V      & 300              & 18.396 &  17.416 & 6  \\
\hline
\end{tabular}
\end{table*}

The telescope was always tracked at sidereal rate. The trailing of the object during the exposure times was negligible. When the time spent between observations made it possible, we aimed the telescope at the same region of the sky each night in order to keep fixed the same stellar field. In other words, the images were not centered on Chiron, which allowed us to always use the same reference stars for all nights in the observing runs in order to minimize systematic photometric errors. Therefore we could perform very high precision relative photometry. The images were corrected for bias frames and twilight flat-field frames that were taken at the beginning of each observation night. The photometric data analysis was accomplished by using differential aperture photometry with our own routines developed in IDL (Interactive Data Language). To obtain Chiron's light curves, we used several reference stars in the field, and tried several synthetic aperture radii in order to chose the aperture that resulted in the best photometry, in terms of lowest scatter (see Table \ref{relative_photometry}).

\begin{table}
\centering
\caption{Parameters used in the photometric analysis. Abbreviations are defined as follows: aperture radius (aper.), radius of the internal annulus for the subtraction of the sky background (an.), width of the subtraction annulus (d$_{\rm an.}$) and  number of reference stars (N$_{\star}$).}
\label{relative_photometry}
\begin{tabular}{ccccc} 
\hline
Date & aper. & an. & d$_{\rm an.}$         & N$_{\star}$    \\
			& (pixels)	 &       (pixels)                &      (pixels)    &      \\
\hline
\hline
2014 Jul 18 & 4 & 12 & 5 &  11\\
2014 Jul 19 & 4 & 12 & 5 & 11\\
2014 Jul 23  & 4 & 12 & 5  &  12\\
2014 Jul 24 & 4 & 12 & 5  &  12\\
2014 Oct 27 & 2   &  9   &  4   &  6\\
2014 Oct 28 & 2   &  9   &  4   &  6 \\
\hline
2015 Sep 11 & 3   &  10  &  3   &  6\\
2015 Sep 13 & 3   &  10  &  3   &  6 \\
2015 Sep 14 & 6  &  9    &  5    &  19$^{\dagger}$\\
\hline
2016 Sep 02 & 6   & 9   &  5   & 1 $^{\ddagger}$\\
\hline
\end{tabular}

\begin{tablenotes}
\small
\item $^{\dagger}$Landolt standard stars: PG2213-006, PG2213-006A, SA110\_229, SA110\_230, SA110\_232, SA110\_233, SA92\_245, SA92\_248, SA92\_249, SA92\_250, SA92\_425, SA92\_426, SA92\_355, SA92\_430, PG2349+002, PG2336-004A, PG2336-004B, SA95\_275 and SA95\_276.\\
\item $^{\ddagger}$Landolt standard star: PG2213\_006.\\
\end{tablenotes}

\end{table}

To determine Chiron's absolute magnitude, we performed an absolute calibration of the images by using Landolt photometric standard stars \citep{Landolt1992}. For the 2015 absolute photometry calibration, the zero point in $R$-band and the extinction coefficient were determined simultaneously using through the Landolt standard stars observed (see footnote at table \ref{relative_photometry}) at several airmasses. On the other hand, for the 2016 absolute photometry calibration, we observed the Landolt standard field PG2213 at several airmasses to check that the night was photometric. The three images of Chiron taken at each filter were directly calibrated using observations of the Landolt standard star  PG2213\_006,
which had been taken at the same airmass as the observations of Chiron; hence, extinction corrections were not needed.

Additionally, we used the collected data to visually search for signs of comet-like activity, like coma or tail, by co-adding the images. The photometric results in form of a rotational light curve, Chiron's absolute brightness measurements and the co-added image are presented and discussed in the next section.

\section{Results and discussion}

\subsection{Rotational light curve}

From the collected images, by using differential aperture photometry we have derived a rotational light curve of Chiron. The photometric measurements were phase plotted to Chiron's known high accuracy synodic rotational period of 5.917813 $ \pm $ 0.000007 h, derived by \cite{MarcialisBuratti1993}.

Because we know that Chiron's light curve is double-peaked \citep{Busetal1989, MarcialisBuratti1993}, we fitted two-term Fourier series to the light curves, and derived light curve amplitudes of 0.016 $ \pm $ 0.004 mag, and 0.020 $ \pm $ 0.005 mag, for the data collected in observation campaigns in 2014 and 2015 respectively. The procedures we followed are describe in \cite{FernandezValenzuelaetal2016}.

\begin{figure}
\begin{center}
\includegraphics[width=\columnwidth]{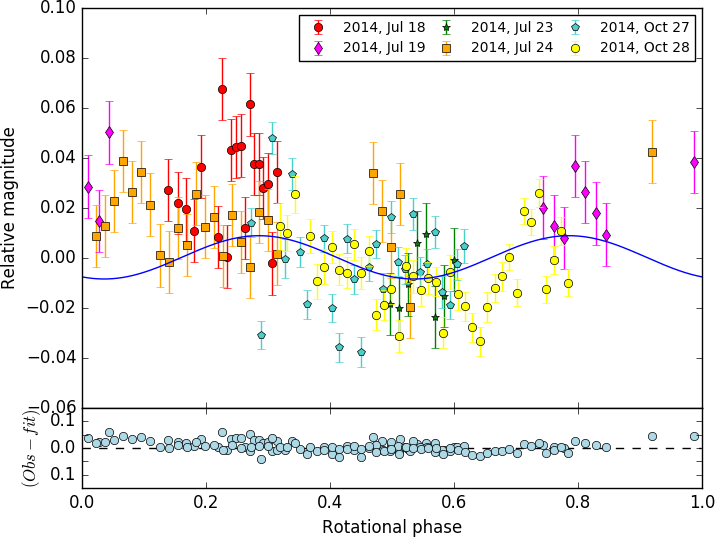}
\caption{The upper plot contains 136 measurements in total, and shows Chiron's rotational light curve resulting from the relative photometry in 2014 with MOSCA 3.5 m (Jul. 18, 19) and DLR-MKIII CCD 1.23 m (Jul. 23, 24 and Oct. 27, 28) instruments at CAHA. The relative magnitude is phase plotted versus Chiron's rotational period of 5.917813 h. A two-term Fourier fit showing a peak-to-peak amplitude of 0.016 $ \pm $ 0.004 mag is superimposed. The Julian date for zero phase angle is 2456857.0 days. The lower plot shows the residuals to the two-term Fourier fit, with a mean value of 0.021 mag.}
\label{chiron_LC_2014}
\end{center}
\end{figure}

\begin{figure}
\begin{center}
\includegraphics[width=\columnwidth]{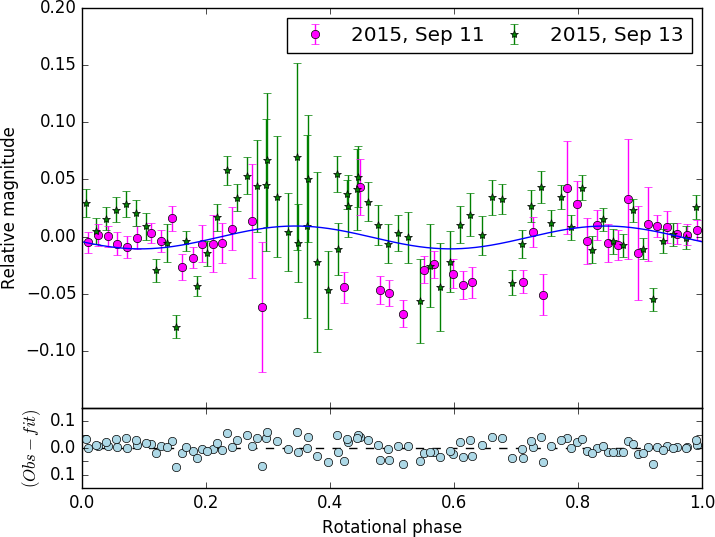}
\caption{The upper plot contains 116 measurements in total, and shows Chiron's rotational light curve resulting from the relative photometry with CAFOS 2.2 m CAHA telescope. The relative magnitude is phase plotted versus Chiron's rotational period of 5.917813 h. A two-term Fourier fit showing a peak-to-peak amplitude of 0.020 $ \pm $ 0.005 mag is superimposed. The Julian date for zero phase angle is 2456857.0 days. The lower plot shows the residuals to the two-term fit, with a mean value of 0.029 mag.}
\label{chiron_LC_Sep11_13_2015}
\end{center}
\end{figure}

The light curves resulting from the relative photometry obtained in the first and second campaign are shown in Figures~\ref{chiron_LC_2014} and ~\ref{chiron_LC_Sep11_13_2015}. 
The upper plots show Chiron's rotational light curve, with a two-term Fourier fit superimposed to the data points. The lower plots show the residuals to the two-term fit.

\begin{figure}
\begin{center}
\includegraphics[angle=0, width=\columnwidth]{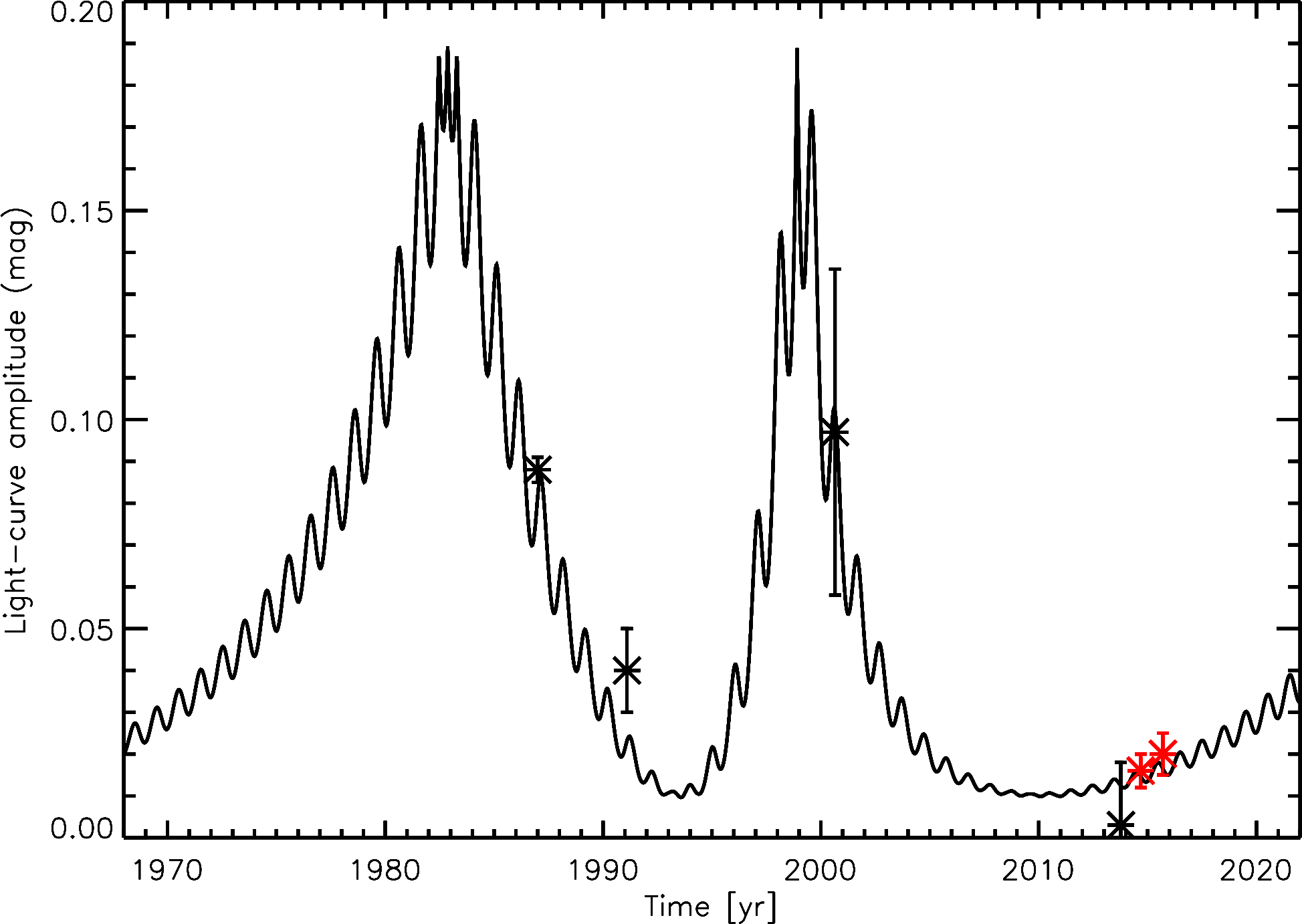}
\caption {The model computed by Ortiz et al. (2015), showing the amplitude of Chiron's light curve versus time. The model takes into account Chiron's triaxial shape, and also the brightness contribution of its rings. The black markers represent observed light curve amplitudes obtained by different observers. The red markers show our new measurements obtained in September 2014 and September 2015.}
\label{chiron_deltamag}
\end{center}
\end{figure}

\cite{Ortizetal2015} have derived Chiron's pole direction from the observed occultations, and modeled the shape of its nucleus as a triaxial ellipsoid. By using axial ratios a/c and b/c of 1.43 and 1.2 respectively, and adding the flux contribution of Chiron's rings, they have computed the decrease of its amplitude as a function of time.
Figure~\ref{chiron_deltamag} shows the modeled amplitude of Chiron's light curve as a function of time, taking into account its triaxial shape, and also the brightness contribution of its rings, computed by \cite{Ortizetal2015}. The black markers represent light curve amplitudes obtained by \cite{Busetal1989}, \cite{Groussinetal2004}, and \cite{Ortizetal2015}. The red markers show light curve amplitudes of 0.016 $ \pm $ 0.004 mag and 0.020 $ \pm $ 0.005 mag, derived from our observations in 2014 and 2015, respectively. As can be seen, the derived light curve amplitudes nicely agrees with the model proposed by \cite{Ortizetal2015}.

\subsection{Absolute brightness}

In order to explain Chiron's absolute magnitude, \cite{Ortizetal2015} showed that using a specific pole orientation, Chiron’s long term brightness variations are compatible with a simple model that incorporates the changing brightness of the rings as the tilt angle with respect to the Earth changes with time. 

To determine Chiron's absolute magnitude from our data set, an absolute calibration of the images was performed by using Landolt photometric standard stars. For the absolute photometry calibration of the data set collected in 2015, the zero point in R and the extinction coefficient were determined simultaneously by observing Landolt stars at several airmasses. Using the derived zero point and the extinction, the apparent magnitude of Chiron was derived in a set of images of Chiron, giving an average value of 17.64 $ \pm $ 0.05 mag(R), which gives an absolute magnitude of 5.17 $ \pm $ 0.05 mag(R). The uncertainty includes the uncertainty in the absolute calibration using Landolt stars together with the uncertainty of the aperture photometry of the Chiron measurements on the night of September 14, 2015. 

This measurement is around 0.4 mag brighter than expected, which can be consistent with the fact that a small tail was detected (discussed in subsection~\ref{tail}). To determine whether Chiron has returned to a normal brightness level, on September 02, 2016, an additional observation was accomplished, showing a still increased absolute magnitude of 5.38 mag(R) and 5.71 mag(V).

\begin{figure}
\begin{center}
\includegraphics[angle=0, width=\columnwidth]{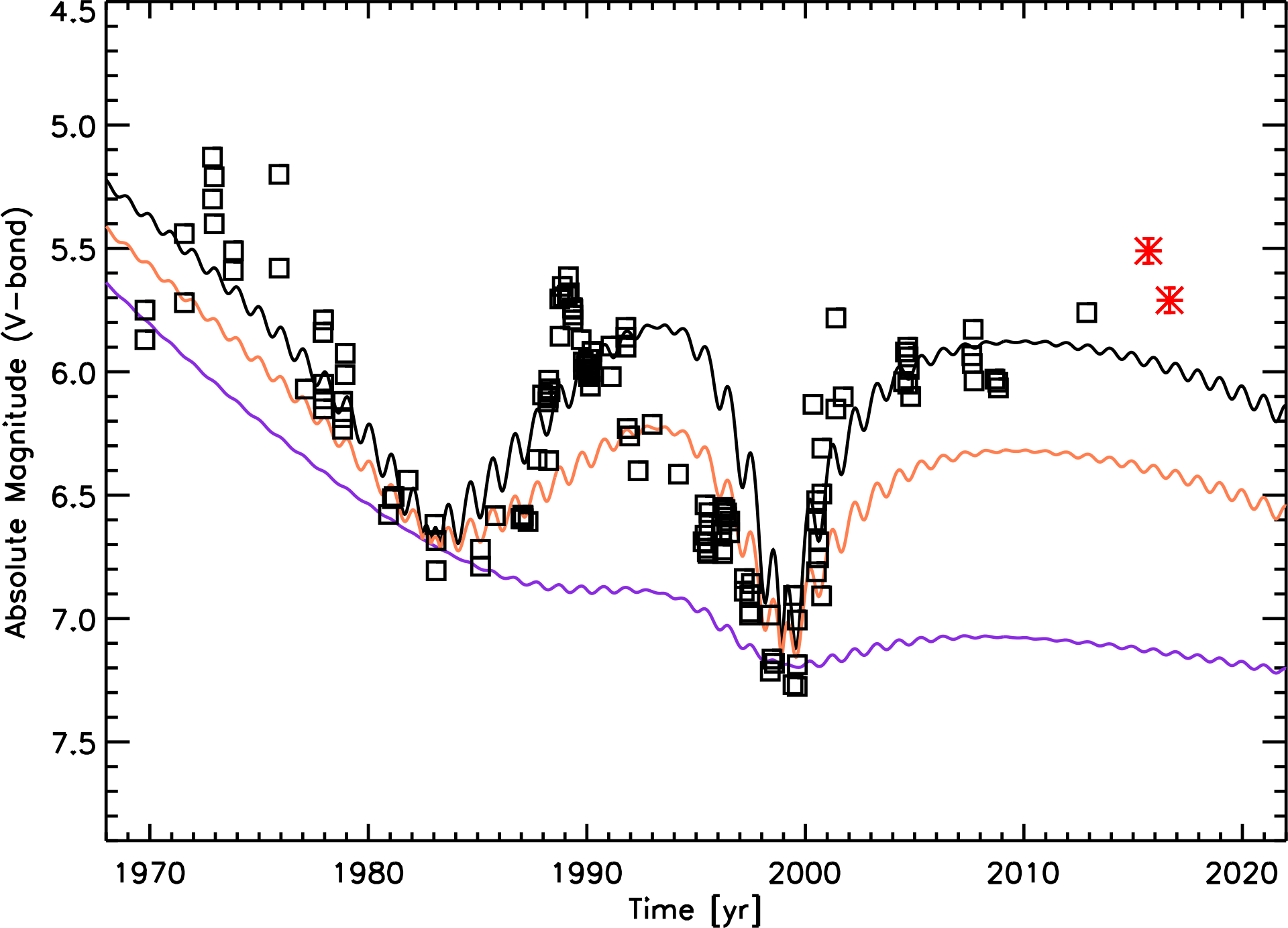}
\caption {Model proposed by Ortiz et al. (2015), that shows the changing brightness of Chiron and the rings depending on Chiron's tilt angle with respect to the Earth, and includes a vanishing coma. 
The model which provides the best agreement with the observations, with the nominal value of the rings' geometric albedo of 0.17 is shown with the black (upper) line. The orange (middle) line corresponds to a model with ring albedo of 0.1, while the purple (lower) line corresponds to a model without a ring contribution at all.
The square symbols represent observations as compiled in Belskaya et al. (2010). The red markers show our measurements of September 2015, and September 2016, obtained in the present work.}
\label{chiron_absmag}
\end{center}
\end{figure}

Figure~\ref{chiron_absmag} shows the model proposed by \cite{Ortizetal2015}, that represents the changing brightness of Chiron's system depending on Chiron's tilt angle with respect to the Earth, and includes the contribution of an exponentially decaying coma. Additionally, the effect of the large relative uncertainty of the geometric albedo of the rings is shown by choosing different albedo values for the ring. The model which provides the best agreement with the observations, with the nominal value of the rings' geometric albedo of 0.17 is shown with the black (upper) line. The orange (middle) line corresponds to a model with ring albedo of 0.1, while the purple (lower) line corresponds to a model without a ring contribution at all.
The square symbols represent observations as compiled in \cite{Belskayaetal2010}. The red markers show our measurements of September 2015, and September 2016, obtained in the present work. The V magnitude for 2015 was obtained by using the V-R colour obtained in 2016, applied to the R measurement from 2015, resulting with an absolute magnitude of 5.51 $ \pm $ 0.05 mag(V).

Although the observed absolute magnitudes of Chiron are slightly brighter than expected, they are in good agreement with the model proposed by \cite{Ortizetal2015}. Therefore, our observations support the presumption that Chiron possesses rings.

\subsection{Photometric scatter and Chiron's microactivity}

An additional motivation for this research was to study whether Chiron has continuous micro-brightness changes that are, according to our hypothesis, possibly caused by debris impacting on its surface. For that reason, while computing photometric light curves of Chiron in our photometric campaigns in 2014 and 2015, we chose a comparison stars of similar brightness as Chiron, and compared the scatter of its photometric measurements to the scatter of Chiron's measurements.

\begin{figure}
\begin{center}
\includegraphics[width=\columnwidth]{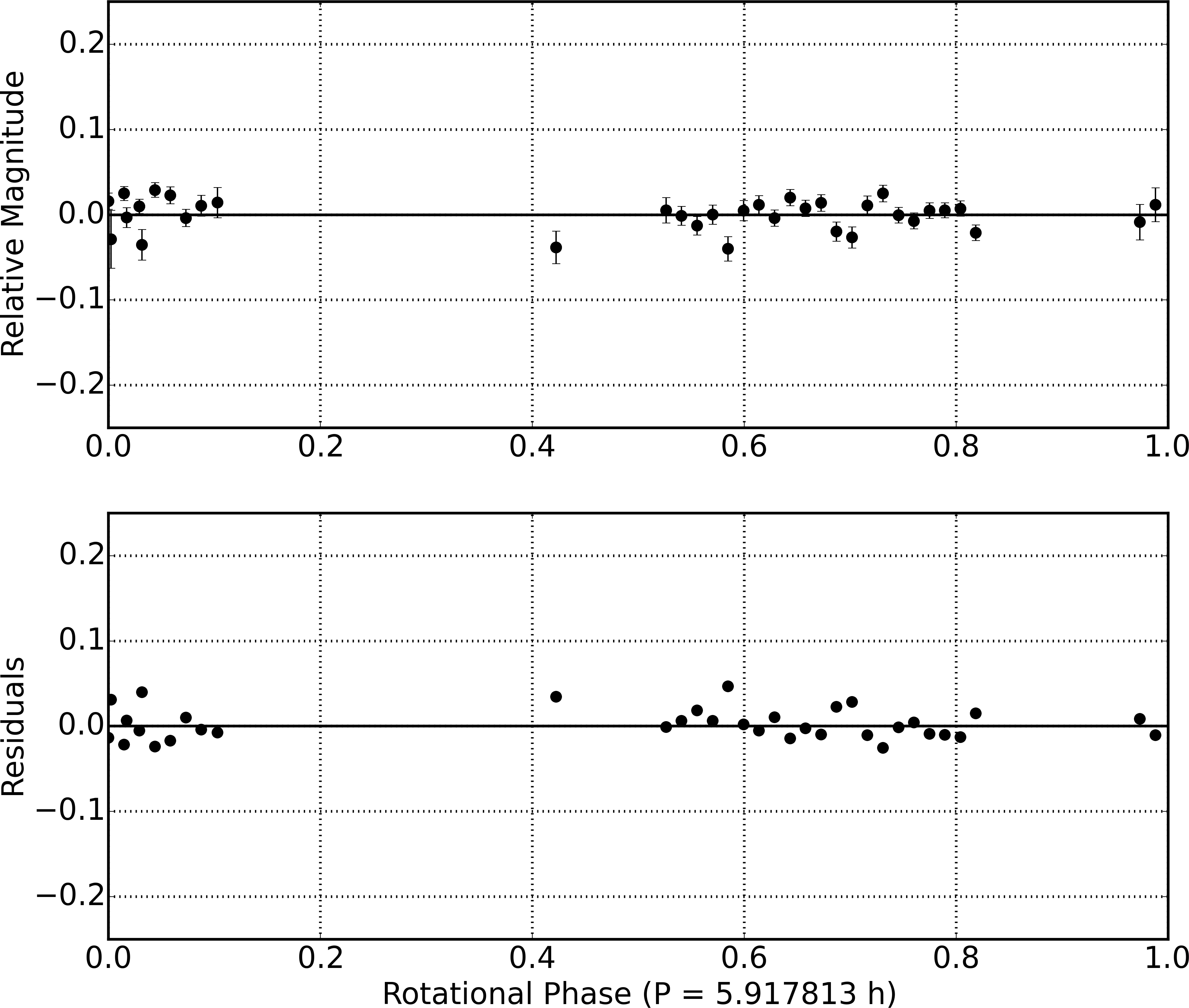}
\caption{The upper plot shows the light curve of a comparison star, resulting from the relative photometry with DLR-MKIII 1.23 m CAHA telescope obtained on July 23 and 24, 2014. The relative magnitude versus time is phase plotted, using Chiron's rotational period. The lower plot shows the residuals, with a mean value of 0.014 mag.}
\label{star_LC_2014}
\end{center}
\end{figure}

\begin{figure}
\begin{center}
\includegraphics[width=\columnwidth]{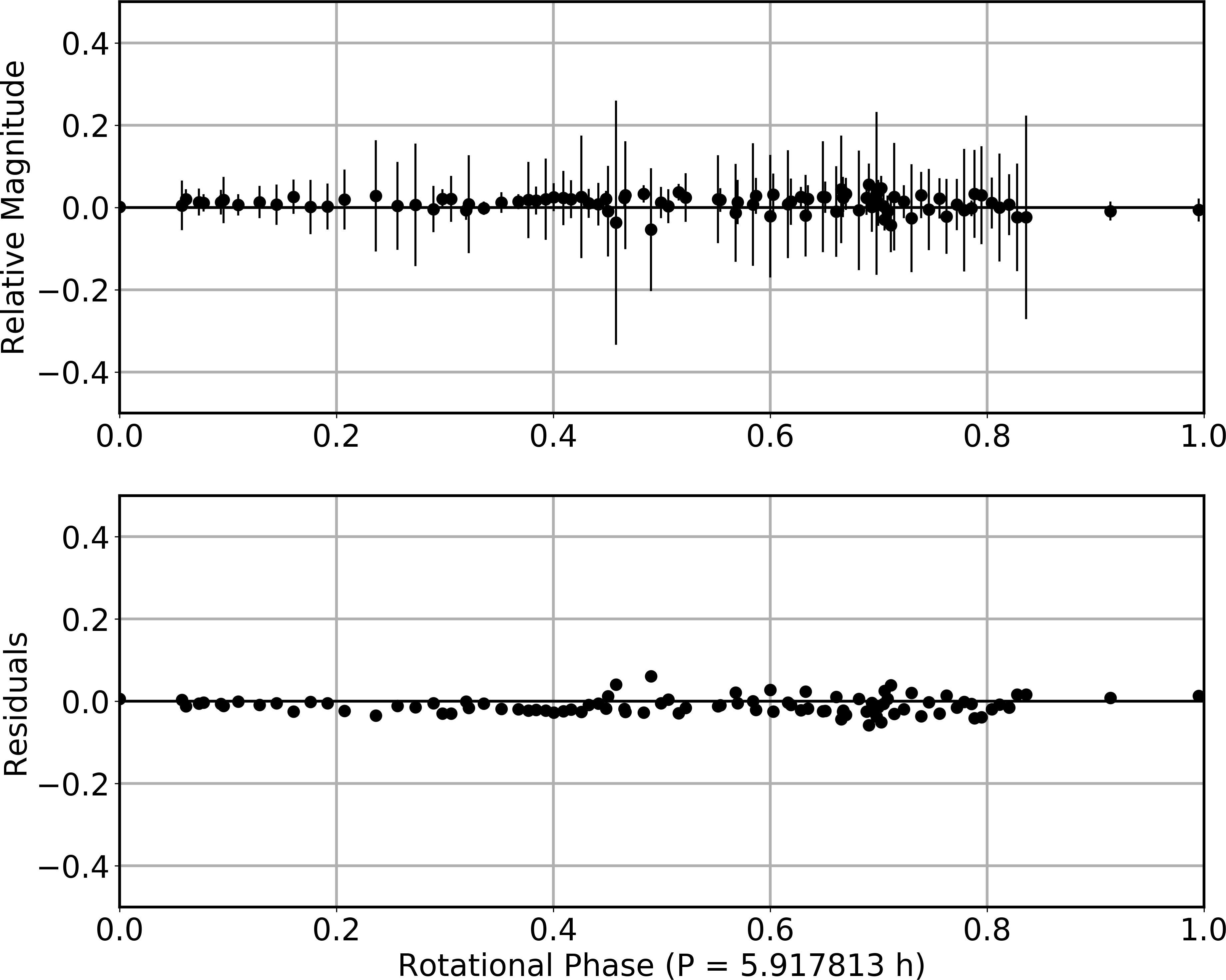}
\caption{The upper plot shows the light curve of a comparison star, resulting from the relative photometry with CAFOS 2.2 m CAHA telescope obtained on September 11-14, 2015. The relative magnitude versus time is phase plotted, using Chiron's rotational period. The lower plot shows the residuals, with a mean value of 0.018 mag.}
\label{star_LC_2015}
\end{center}
\end{figure}

The light curves of the comparison stars are shown in Figures~\ref{star_LC_2014} and ~\ref{star_LC_2015} for the photometric campaigns in 2014 and 2015, respectively. Because Chiron has a determined rotational period, the comparison stars' light curves were also phase plotted, using Chiron's accurate rotational period.
Our campaign in 2014 contains three runs. To study photometric scatter, we chose a comparison star from the run carried out on July 23 and 24, 2014, with DLR-MKIII 1.23 m CAHA telescope. The scatter of Chiron's photometric measurements for this run values 0.027 mag, while the comparison star's scatter is 0.014 mag.
During the photometric campaign in 2015, containing data collected with CAFOS 2.2 m CAHA telescope on September 11 and September 13, we measured a scatter of Chiron's photometric measurements of 0.029 mag, while the comparison star's scatter values 0.018 mag. 
The noticeable difference between the comparison star's and Chiron's photometric scatter supports our hypothesis of a continuous microactivity on Chiron. 


The microactivity on Chiron might be triggered by any yet unknown mass-loss mechanism. In our hypothesis of microactivity on Chiron in form of small outbursts, we suggest the possibility of debris orbiting around Chiron, especially material diverging from the ring, falling back and impacting on Chiron's surface, producing outbursts and forming a bound or quasi-bound coma. That mechanism could also be a plausible explanation for impulsive brightenings on a timescale of hours that have been observed by \cite{LuuJewitt1990}.
Unlike to a regular cometary coma, which is not gravitationally bound and continually escapes, Chiron's quasi-bound coma would exponentially disappear with a lifetime of several years.

The initial origin of Chiron's coma could indeed be impact-related. \cite{Ortizetal2015} suggested a model of a exponentially decaying coma. Such behavior of a decaying coma has already been noticed by \cite{MeechBelton1990} and \cite{Duffardetal2002}, but \cite{Ortizetal2015} note that the absolute maximum in the photometry happened around 1973, setting that date for the start of the decay.
Because the absolute photometric data from 1970 shows lower values than the maximum around 1973, we think that a major impact could have happened between 1970 and 1973.

The surface brightness measurements of Chiron's coma in December 1989 estimated to 24.8 mag(R)/arcsec$^{2}$ \citep{MeechBelton1990}, and in April 1998 estimated to 26.0 mag(R)/arcsec$^{2}$ (26.2 mag(V)/arcsec$^{2}$)\citep{SilvaCellone2001}, show that the coma was slowly extinguishing.

So far the activity outbursts on regular comets was thought to be due to mass-loss mechanisms such as sublimation of water-ice and thermal processes. But we hypothesize that they could also be triggered by collisions, either with debris around them or with meteoroids.

\subsection{Comet-like tail} \label{tail}

\begin{figure}
\begin{center}
\includegraphics[width=\columnwidth]{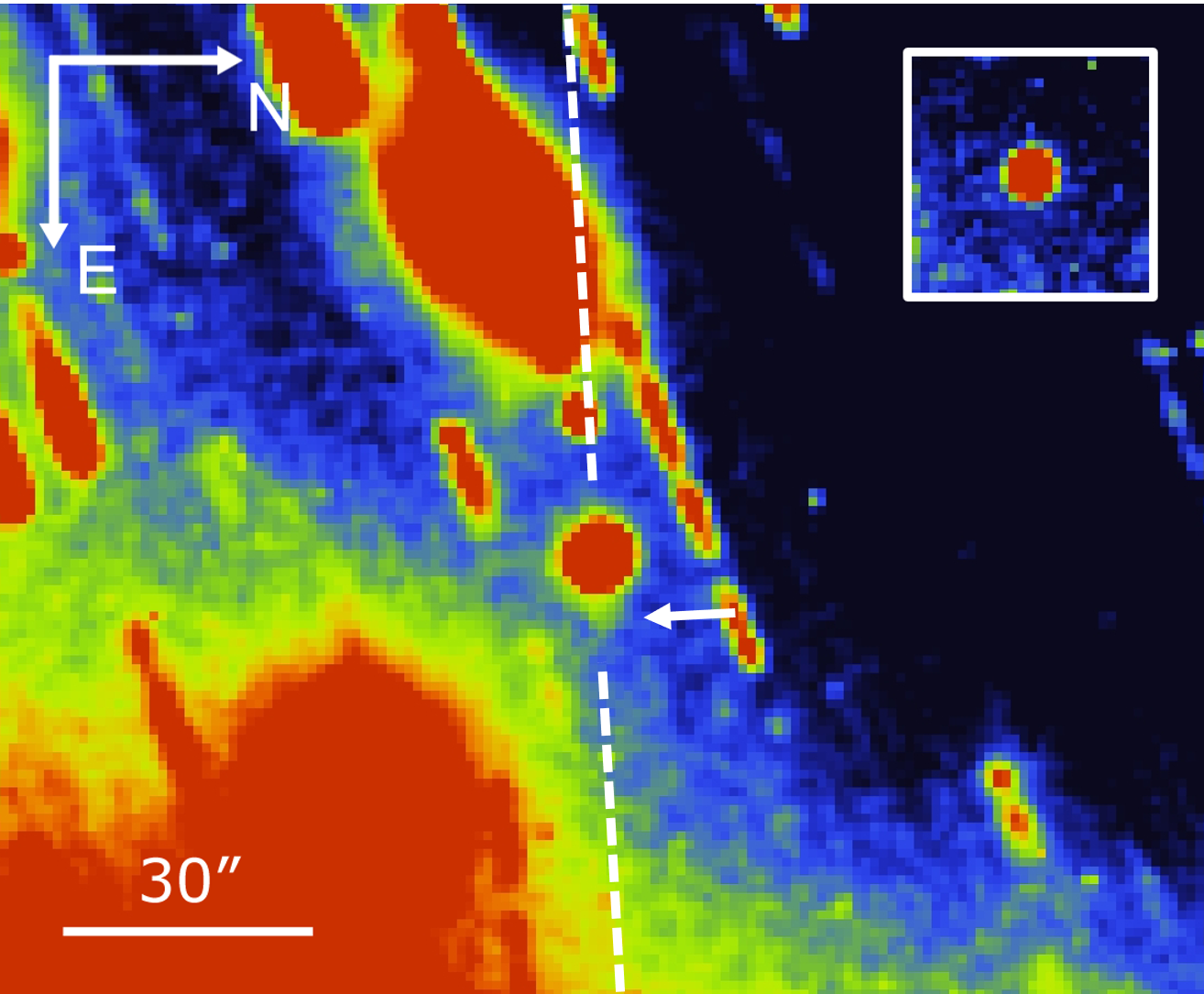}
\caption{A false-colour image of Chiron (in the centre), showing an asymmetric shape, possibly representing a $ \sim $5 arcsec long tail (marked with an arrow) with a measured position angle of $ \sim $87$^{\circ}$ represented by the dashed line. The image was taken with the CAFOS 2.2 m CAHA telescope over 3 nights between September 11 and September 14, 2015, with a total integration time of 28500 sec (475 min). The Julian Date for midpoint of the exposure is 2457278.62984 days. The sub-image in the upper right corner shows a combined false-colour image of an isolated star, co-added with the very same images we used for Chiron's image.}
\label{chiron_tail}
\end{center}
\end{figure}

Additionally, we used the data collected in 2015 to search for comet-like activity signs like coma or tail. Figure~\ref{chiron_tail} shows a combined false-colour image of Chiron that contains 95 integrations, 300 sec each. It was taken with the CAFOS 2.2 m CAHA telescope over 3 nights between September 11 and September 14, 2015, resulting with a total integration time of 28500 sec (475 min). The Julian Date for midpoint of the exposure is 2457278.62984 days (September 13 2015, 03:06:58 UT).

The background stars are trailed because images were aligned to Chiron. The combined image of Chiron shows that the coma itself is not anymore detectable, but a very faint asymmetric shape (marked with an arrow) with a measured position angle of $ \sim $87$^{\circ}$ (represented by the dashed line), that can be explained by a $ \sim $5 arcsec long tail , of a determined surface brightness of 25.3 mag(R)/arcsec$^{2}$, is observed. Because of it's low surface brightness, although the quality of individual images has been carefully examined before co-adding, a contamination from a faint stellar source in the background can not be completely discarded.

The sub-image in the upper right corner of Figure~\ref{chiron_tail} shows a combined false-colour image of an isolated star, co-added with the very same images we used for Chiron's image, but aligned to the stars. We can notice that the contour plots of a co-added star doesn't show any asymmetries similar to those of Chiron, which excludes the possibility that the tail emerged due to a tracking problem in the images.

\begin{figure}
\begin{center}
\includegraphics[width=\columnwidth]{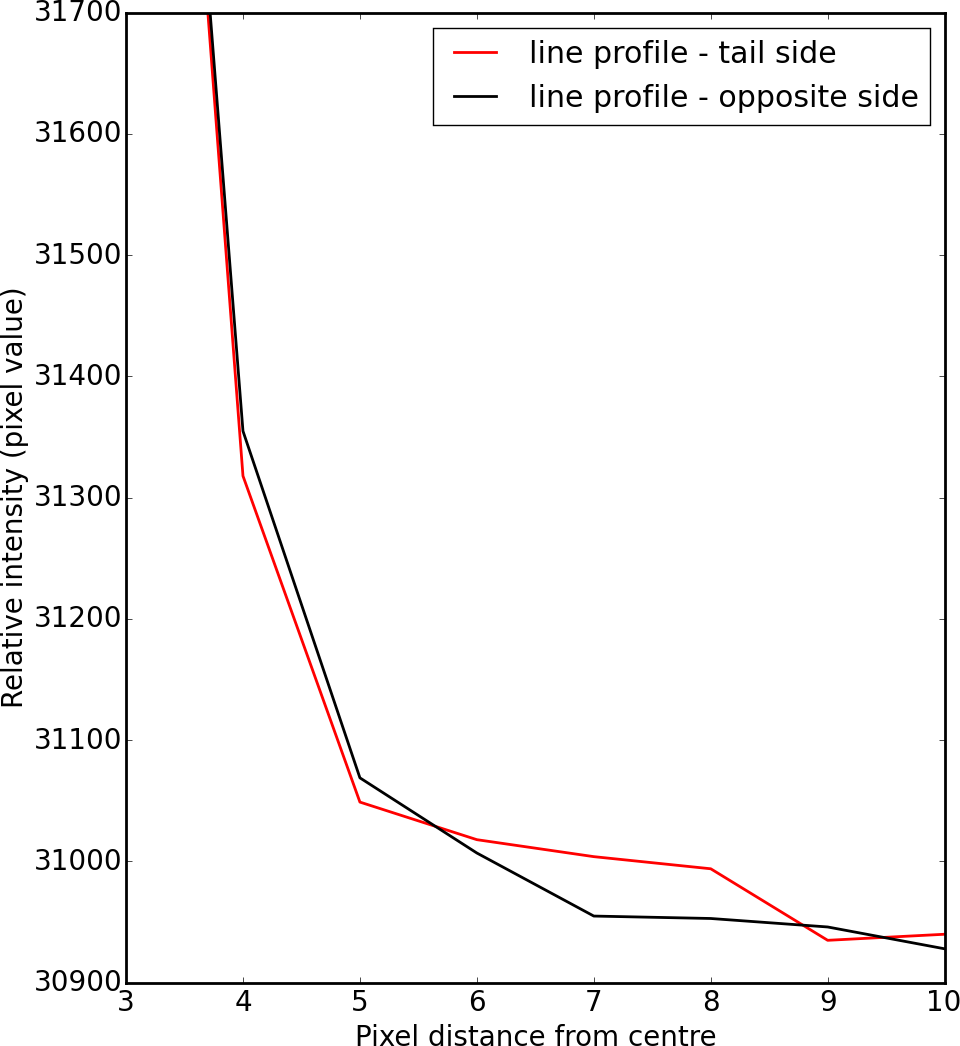}
\caption{Profile line of Chiron showing the asymmetry of a radial cut in the position angle of the tail. The red line shows the relative intensity versus pixel distance from Chiron's centre on the side of the tail. For comparison, the black line shows the relative intensity versus pixel distance from Chiron's centre on the opposite side of the tail. Relative difference in the intensity between the profile lines is about 50.}
\label{chiron_profile}
\end{center}
\end{figure}

The profile line of Chiron is showing an asymmetry of its radial cut in the direction of the position angle of the possible tail. The red line in Figure~\ref{chiron_profile} shows relative intensity versus pixel distance from Chiron's centre on the side of the tail. The black line shows relative intensity versus pixel distance from Chiron's centre on the opposite side of the tail. The relative brightness of Chiron's tail (relative intensity difference of the two lines) is $ \sim $50.

\section{Conclusions}

Although the collision rates in the Centaur region are expected to be very low, and impacts are thought not to be responsible for the mass-loss, since the recent indications that Chiron might possess a ring \citep{Ortizetal2015}, the impact triggered dust release mechanism should not be excluded as a possibility. Assuming that there is debris around Chiron, it is well possible that some of this debris may be continually falling and impacting on the surface. 

From observations collected in campaigns in 2014 and 2015, we found that the photometric scatter in Chiron's data is larger than the control stars' scatter, indicating a microactivity on Chiron, possibly caused by debris and material diverging from Chiron's ring, falling back to its surface, producing outbursts, and forming a bound or quasi-bound coma.

In our visual examination of a combined false colour image of Chiron, containing 95 integrations of 300 sec each, a coma was not detected, but we have found a feature that looks like a $ \sim $5 arcsec long tail, showing a surface brightness of 25.3 mag/arcsec$^{2}$.

The amplitudes of Chiron's rotational light curves derived from observations collected in 2014 and 2014 are in very good agreement with the model proposed by \cite{Ortizetal2015}, assuming that Chiron is a triaxial ellipsoid and taking into account the flux contribution of its ring.

The observed absolute magnitudes of Chiron in 2015 and 2016 are in good agreement with the model that incorporates the changing brightness of the rings depending on Chiron's tilt angle with respect to the Earth, proposed by \cite{Ortizetal2015}. Hence, our observations are consistent with the presumption that Chiron possesses rings. 

The initial origin of Chiron's coma could also be impact-related. Based on models of a exponentially decaying coma, starting its decay between 1970 and 1973 \cite{Ortizetal2015}, the impact could have happened in that period.

We hypothesize that some activity outbursts on regular comets could also be triggered by collisions, either with debris orbiting them, or with meteoroids.

\section*{Acknowledgements}
We are grateful to the CAHA staff. This research is based on observations collected at Centro Astron\'omico Hispano Alem\'an (CAHA) at Calar Alto, operated jointly by the Max-Planck Institut fur Astronomie and the Instituto de Astrof\'isica de Andaluc\'ia (CSIC). The research leading to these results has received funding from the European Union’s Horizon 2020 Research and Innovation Programme, under Grant
Agreement No. 687378. Funding from Spanish grants AYA-2011-30106-CO2-O1 and AYA-2014-56637-C2-1-P are acknowledged, as is the Proyecto de Excelencia de la Junta de Andaluc\'ia, J. A. FEDER funds are also acknowledged.


\bsp	
\label{lastpage}

\begin{thebibliography}{}
\makeatletter


\bibitem[Braga-Ribas et al.(2014)]{BragaRibasetal2014} Braga-Ribas, F., Sicardy, B., Ortiz, J.~L., et al.\ 2014, \nat, 508, 72 

\bibitem[Belskaya et al.(2010)]{Belskayaetal2010} Belskaya, I.~N., Bagnulo, S., Barucci, M.~A., et al.\ 2010, \icarus, 210, 472 

\bibitem[Bus et al.(1989)]{Busetal1989} Bus, S.~J., Bowell, E., Harris, A.~W., \& Hewitt, A.~V.\ 1989, \icarus, 77, 223 

\bibitem[Bus et al.(1991)]{Busetal1991} Bus, S.~J., A'Hearn, M.~F., Schleicher, D.~G., \& Bowell, E.\ 1991, Science, 251, 774 

\bibitem[Bus et al.(1996)]{Busetal1996} Bus, S.~J., Buie, M.~W., Schleicher, D.~G., et al.\ 1996, \icarus, 123, 478 

\bibitem[Bus et al.(2001)]{Busetal2001} Bus, S.~J., A'Hearn, M.~F., Bowell, E., \& Stern, S.~A.\ 2001, \icarus, 150, 94 

\bibitem[Duffard et al.(2002)]{Duffardetal2002} Duffard, R., Lazzaro, D., Pinto, S., et al.\ 2002, \icarus, 160, 44 

\bibitem[Duffard et al.(2014)]{Duffardetal2014} Duffard, R., Pinilla-Alonso, N., Ortiz, J.~L., et al.\ 2014, \aap, 568, A79 

\bibitem[Durda \& Stern(2000)]{DurdaStern2000} Durda, D.~D., \& Stern, S.~A.\ 2000, \icarus, 145, 220 

\bibitem[Elliot et al.(1995)]{Elliotetal1995} Elliot, J.~L., Olkin, C.~B., Dunham, E.~W., et al.\ 1995, \nat, 373, 46 

\bibitem[Fern{\'a}ndez-Valenzuela et al.(2016)]{FernandezValenzuelaetal2016} Fern{\'a}ndez-Valenzuela, E., Ortiz, J.~L., Duffard, R., Santos-Sanz, P., \& Morales, N.\ 2016, \mnras, 456, 2354 

\bibitem[Fornasier et al.(2013)]{Fornasieretal2013} Fornasier, S., Lellouch, E., M{\"u}ller, T., et al.\ 2013, \aap, 555, A15 

\bibitem[Groussin et al.(2004)]{Groussinetal2004} Groussin, O., Lamy, P., \& Jorda, L.\ 2004, \aap, 413, 1163 

\bibitem[Hahn \& Bailey(1990)]{HahnBailey1990} Hahn, G., \& Bailey, M.~E.\ 1990, \nat, 348, 132 

\bibitem[Hartmann et al.(1990)]{Hartmannetal1990} Hartmann, W.~K., Tholen, D.~J., Meech, K.~J., \& Cruikshank, D.~P.\ 1990, \icarus, 83, 1 

\bibitem[Jewitt(2012)]{Jewitt2012} Jewitt, D.\ 2012, \aj, 143, 66 

\bibitem[Jewitt et al.(2011)]{Jewitt2011} Jewitt, D., Weaver, H., Mutchler, M., Larson, S., \& Agarwal, J.\ 2011, \apjl, 733, L4 

\bibitem[Kowal et al.(1979)]{Kowaletal1979} Kowal, C.~T., Liller, W., \& Marsden, B.~G.\ 1979, Dynamics of the Solar System, 81, 245 

\bibitem[Landolt(1992)]{Landolt1992} Landolt, A.~U.\ 1992, \aj, 104, 340 

\bibitem[Luu et al.(2000)]{Luuetal2000} Luu, J.~X., Jewitt, D.~C., \& Trujillo, C.\ 2000, \apjl, 531, L151 

\bibitem[Luu \& Jewitt(1990)]{LuuJewitt1990} Luu, J.~X., \& Jewitt, D.~C.\ 1990, \aj, 100, 913 

\bibitem[Marcialis \& Buratti(1993)]{MarcialisBuratti1993} Marcialis, R.~L., \& Buratti, B.~J.\ 1993, \icarus, 104, 234 

\bibitem[Meech \& Belton(1990)]{MeechBelton1990} Meech, K.~J., \& Belton, M.~J.~S.\ 1990, \aj, 100, 1323 

\bibitem[Moreno et al.(2011)]{Moreno596} Moreno, F., Licandro, J., Ortiz, J.~L., et al.\ 2011b, \apj, 738, 130

\bibitem[Moreno et al.(2011)]{MorenoA2} Moreno, F., Licandro, J., Tozzi, G.~P., et al.\ 2011a, Highlights of Spanish Astrophysics VI, 587 

\bibitem[Oikawa \& Everhart(1979)]{OikawaEverhart1979} Oikawa, S., \& Everhart, E.\ 1979, \aj, 84, 134 

\bibitem[Ortiz et al.(2015)]{Ortizetal2015} Ortiz, J.~L., Duffard, R., Pinilla-Alonso, N., et al.\ 2015, \aap, 576, A18 

\bibitem[Romon-Martin et al.(2003)]{RomonMartinetal2003} Romon-Martin, J., Delahodde, C., Barucci, M.~A., de Bergh, C., \& Peixinho, N.\ 2003, \aap, 400, 369 

\bibitem[Ruprecht et al.(2015)]{Ruprechtetal2015} Ruprecht, J.~D., Bosh, A.~S., Person, M.~J., et al.\ 2015, \icarus, 252, 271 

\bibitem[Sarid \& Prialnik(2009)]{SaridPrialnik2009} Sarid, G., \& Prialnik, D.\ 2009, Meteoritics and Planetary Science, 44, 1905 

\bibitem[Silva \& Cellone(2001)]{SilvaCellone2001} Silva, A.~M., \& Cellone, S.~A.\ 2001, \planss, 49, 1325 

\bibitem[Snodgrass et al.(2010)]{Snodgrass2010} Snodgrass, C., Tubiana, C., Vincent, J.-B., et al.\ 2010, \nat, 467, 814

\bibitem[Tedesco(1979)]{Tedesco1979} Tedesco, E.~F.\ 1979, \icarus, 40, 375 

\bibitem[Tholen et al.(1988)]{Tholenetal1988} Tholen, D.~J., Hartmann, 
W.~K., Cruikshank, D.~P., et al.\ 1988, \iaucirc, 4554, 2 

\bibitem[Zappal{\`a} et al.(2002)]{Zappalaetal2002} Zappal{\`a}, V., Cellino, A., \& Dell'Oro, A.\ 2002, \icarus, 157, 280 





\makeatother
\end{thebibliography}
\end{document}